\documentclass[aps,superscriptaddress,twocolumn,longbibliography]{revtex4-2}
\usepackage{graphicx,xcolor}
\usepackage{bm,bbm,amsmath,amssymb}
\usepackage{physics}
\usepackage{dsfont}

\usepackage[urlcolor=purple,colorlinks=true,citecolor=blue,	linkcolor=red,pdfstartview={FitH},bookmarks=false]{hyperref}
\DeclareMathAlphabet\mathbfcal{OMS}{cmsy}{b}{n}
\usepackage{cancel}

\usepackage{mathtools}

\begin{document}

\title{Surviving readout errors and incomplete postselection in monitored quantum circuits}
\author{Murtaza Ali Mir}
\affiliation{Computational Physics Laboratory, Physics Unit, Faculty of Engineering and Natural Sciences, Tampere University, P.O. Box 692, FI-33014 Tampere, Finland}

\author{Teemu Ojanen}
\affiliation{Computational Physics Laboratory, Physics Unit, Faculty of Engineering and Natural Sciences, Tampere University, P.O. Box 692, FI-33014 Tampere, Finland}
\affiliation{Helsinki Institute of Physics P.O. Box 64, FI-00014, Finland}

\author{Ali G.  Moghaddam}\email{Email: ali.moghaddam@aalto.fi}
\affiliation{Department of Applied Physics, Aalto University, 02150 Espoo, Finland}
\affiliation{Computational Physics Laboratory, Physics Unit, Faculty of Engineering and Natural Sciences, Tampere University, P.O. Box 692, FI-33014 Tampere, Finland}
\affiliation{Helsinki Institute of Physics P.O. Box 64, FI-00014, Finland}

\begin{abstract}
We study how readout errors affect dynamics in monitored quantum circuits. Such errors make postselection effectively incomplete, conditioning trajectories on imperfect rather than true records. Focusing on U(1)-symmetric circuits, we find that the charge-sharpening transition survives at a finite error rate, with the sharp phase's vanishing fluctuations replaced by a finite steady-state plateau. The charge fluctuations show a shallow minimum near the ideal transition point as a function of measurement rate, signaling the competition between measurement-induced purification and noise-induced mixing. By decomposing the fluctuations into quantum and classical contributions, we show the minimum marks where incomplete postselection still retains enough trajectory memory to attenuate noise-induced mixing, leaving a detectable signature of the transition even in a noisy setting. Mid-circuit measurements are far harder to mitigate against readout errors than standard end-of-circuit ones, because the errors corrupt the postselection itself; the decomposition therefore offers a minimal mitigation framework and a route to locate the transition directly from experimental data. Our results show that charge sharpening survives readout errors in a quantitative and experimentally accessible way.
\end{abstract}

\maketitle

\emph{\color{blue} Introduction}.---Understanding the relationship between physical dynamics and quantum information in complex, far-from-equilibrium systems has become a central theme in the study of quantum many-body physics. A particularly fertile arena, termed \emph{monitored dynamics}, has emerged in which the competition between unitary evolution and non-unitary monitoring by local measurements gives rise to rich phenomena organized around \emph{measurement-induced phase transitions} (MIPTs) \cite{li_quantum_2018,skinner_measurement-induced_2019,potter_entanglement_2021,Fisher_random_quantum_circuits,bao_theory_2020,li_measurement-driven_2019,Jian-Ludwig2020_measurement_criticality}. In this paradigm, many-body unitary dynamics drives entanglement growth and information scrambling, while measurement suppresses and restructures entanglement, so that their competition produces sharp transitions in the entanglement structure and associated physical properties of the system.
While entanglement dynamics has been central to this picture, a diverse range of related transitions has also been identified, including purification, charge-sharpening, and transitions in magic (non-stabilizerness), among others \cite{gullans_dynamical_2020,Agrawal_2022,Halpern2023,Gullans2024magic,Zarand2025,choi_quantum_2020,Yamamoto2023}.
These systems pose genuinely hard problems in physics: the interplay of symmetry, measurement, and far-from-equilibrium unitary dynamics generates a landscape in which entanglement structure, emergent quantum order, and thermalization compete in ways that remain only partially understood \cite{
PRXQuantum.4.040309,ippoliti_entanglement_2021,nahum_measurement_2021,vasseur2026houches,Pal_2020_MBL_MIPT,Mirlin2023,Marino2025_symmetry,Michailidis20revivals,Turkeshi2022_data,Kawabata2026_symmetry}.

Beyond their intrinsic theoretical interest, monitored symmetric circuits occupy a privileged position as platforms for probing the capabilities of present-day quantum hardware to the extreme. Unlike standard quantum algorithms or Hamiltonian dynamics, they require mid-circuit measurements and are consequently subject to the associated postselection problem \cite{Ippoliti_postselection,McGinley2024,XEB-altman,poyhonen2024,Fazio2024postselection}; they further involve a highly nontrivial interplay between measurement, symmetry, and information flow, both of which make them difficult to realize on NISQ-era devices. Despite this, several experiments with modest qubit counts and shallow circuits have been attempted~\cite{Monroe_measurement_2022,google2023measurement,Minnich_2023measurement,XEB-experiment}, though imperfections such as finite coherence times and unavoidable local noise, particularly readout errors \cite{mitigation_RMP,Gambetta2021ScalableM3,Nachman2020Unfolding,Wei2019imperfect_measurement,Maciejewski2021modelingmitigation,Wei2019imperfect_measurement,Nachman_activereadout}, make the study of monitored dynamics on these devices considerably more challenging. Monitored symmetric circuits thus serve a dual purpose: as a practical testbed for benchmarking and characterizing NISQ devices in physically meaningful settings, and as a source of fundamental questions about entanglement, conserved-charge dynamics, quantum criticality, and the nature of information in monitored many-body systems.

\begin{figure}
    \centering
    \includegraphics[width=0.8\linewidth]{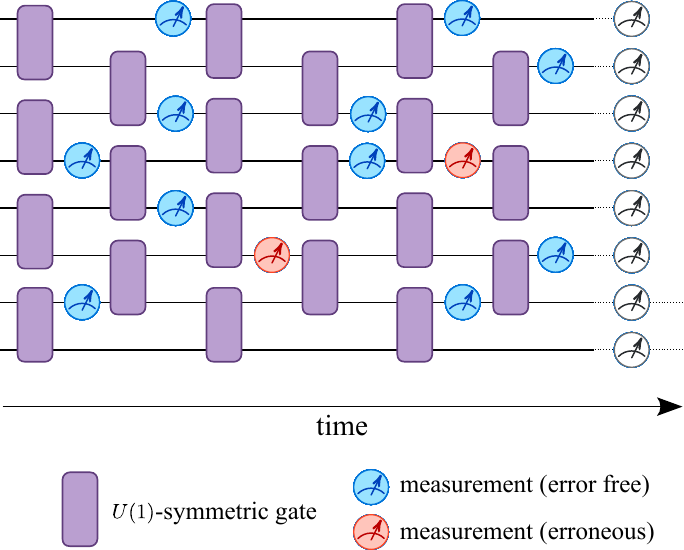}
    \caption{Schematic of a U(1)-symmetric monitored circuit. Two-qubit gates (purple blocks) are drawn from the Haar measure restricted to U(1) charge-conserving unitaries. After each layer of gates, every qubit is measured independently with probability $p_m$. Correctly recorded measurement outcomes are shown in blue, and outcomes affected by assignment (readout) errors are shown in red.}
    \label{fig:1}
\end{figure}

In this letter, we investigate the robustness of \emph{charge-sharpening dynamics} \cite{Agrawal_2022,Chakraborty2024} in U(1)-symmetric monitored quantum circuits in the presence of readout errors (a schematic of the circuit is illustrated in Fig. \ref{fig:1}). Because postselection is based on reported rather than true outcomes, assignment errors in readout effectively misclassify measurement records, giving rise to a incomplete or \emph{partial postselection} despite nominal conditioning on specific outcomes. We show that such errors induce a systematic broadening of the charge distribution within postselected ensembles: the ideal charge-sharpening phase persists at moderate error rates, but with a steady state exhibiting small, finite charge fluctuations rather than a sharply defined charge sector. Analyzing these steady-state fluctuations as a function of measurement rate reveals scaling behavior that captures the competition between measurement-induced purification and \emph{noise-induced mixing}, with the effective broadening set by the assignment error rate and suppressed at higher measurement rates. Charge sharpening is thus observable even in the presence of realistic readout errors, both because moderate noise broadens rather than destroys the transition and because the residual fluctuations scale predictably with measurement rate; when assignment errors dominate, these fluctuations directly diagnose the underlying noise rate, pointing to a practical route for estimating and mitigating readout errors in monitored quantum systems.

\emph{\color{blue} Charge-sharpening transition}.---Monitored quantum circuits are built from an alternating sequence of local unitary gates and projective measurements. When the unitaries are chosen to conserve a global $\text{U(1)}$ charge, the resulting circuit dynamics respects this symmetry while measurements probe and progressively reduce the associated charge fluctuations. In such circuits, the system undergoes a \emph{charge-sharpening transition} that is distinct from, and typically precedes, the entanglement and purification transitions ($p_\# < p_c$). Starting from a mixed state or a state prepared as a superposition of many charge sectors, this transition marks the point at which the system becomes effectively confined to a single, fixed-charge Hilbert subspace, suppressing the ``mixedness'' associated with fluctuations between charge sectors. Crucially, while this phase collapses the distribution across sectors, it does not imply total purification of the state: significant mixedness can still persist within the fixed-charge sector itself. A system may therefore reside in a charge-sharp phase with a volume-law entangled, non-purified structure, with the subsequent transition to a fully purified, low-entanglement state occurring only at a higher measurement rate.
\par
Since the resulting sharp charge sector is determined by the specific sequence of reported measurement outcomes, and different outcome sequences generally sharpen to different charge sectors, charge sharpening is inherently more sensitive to readout errors than the entanglement or purification transitions. A single misassigned outcome causes the inferred charge sector to diverge from the true one, effectively mixing distinct charge sectors under postselection. This is precisely why, in practice, conditioning on reported outcomes amounts to only partial postselection: the ensemble nominally restricted to a fixed charge sector in fact retains a residual admixture of others, as detailed below.

\emph{\color{blue}Readout noise}.---In the presence of assignment errors, the reported measurement outcome can differ from the true one. For example, a true outcome $0$ may be misreported as $1$, and conversely a true outcome $1$ may be misreported as $0$. The probability of observing a reported outcome $\tilde{m}_i$ is then given by
\begin{align}\label{eq:stochastic-assignemnt}
\begin{pmatrix}
\tilde{\pi}_{0} \\
\tilde{\pi}_{1}
\end{pmatrix}
= \Gamma \begin{pmatrix}
\pi_{0} \\
\pi_{1}
\end{pmatrix}
=
\begin{pmatrix}
1-\eta & \eta' \\
\eta & 1-\eta'
\end{pmatrix}
\begin{pmatrix}
\pi_{0} \\
\pi_{1}
\end{pmatrix},
\end{align}
where $\pi_{m}$ are the ideal Born probabilities of the true outcome $m_i$, and $\tilde\pi_{\tilde m}$ those of the reported outcome $\tilde m_i$. We allow for asymmetric assignment errors, with $\eta$ the probability of misassigning $0$ as $1$ and $\eta'$ the probability of misassigning $1$ as $0$. The matrix elements $\Gamma_{\tilde m m}=p_{\tilde m|m}$ are the conditional probabilities of reporting $\tilde m$ given the true outcome $m$. Hence, $\Gamma$ must be left-stochastic, i.e., each column sums to one, $\sum_{\tilde m=0}^{1} p_{\tilde m|m}=1$.

The presence of assignment errors mixes the probabilities of different ideal measurement outcome vectors. Postselecting on a given noisy outcome sequence $\tilde{\bf m}$ therefore yields a mixed state, formed from the ideal post-measurement states corresponding to different ideal outcome vectors $\mathbf{m}$, weighted by the probability of reporting $\tilde{\mathbf{m}}$ given $\mathbf{m}$ together with the probability of $\mathbf{m}$ itself:
\begin{align}\label{eq:noisy_ensemble}
    \rho_{\tilde{\mathbf{m}}} = \frac{\sum_{\mathbf{m}} P_{\tilde{\mathbf{m}}|\mathbf{m}} P_{\mathbf{m}} \ket{\psi_{\mathbf{m}}}\bra{\psi_{\mathbf{m}}}}
    {\sum_{\mathbf{m}} P_{\tilde{\mathbf{m}}|\mathbf{m}} P_{\mathbf{m}}},
\end{align}
where $\ket{\psi_{\mathbf{m}}}$ is the post-measurement state corresponding to the ideal outcome vector $\mathbf{m}$. Denoting by $P_{\bf m}$ the probability of obtaining the ideal outcome sequence ${\bf m}=(m_1,\cdots,m_M)$, the probability of observing a given reported outcome vector $\tilde{\mathbf{m}}$ in the presence of assignment errors is $\tilde{P}_{\tilde{\mathbf{m}}}= \sum_{\mathbf{m}} {P}_{\tilde{\mathbf{m}}|\mathbf{m}}\: {P}_{\mathbf{m}}$.
In the two extreme limits of perfect measurement (no assignment errors) and completely random assignment (maximal assignment errors, $\eta=\eta'=1/2$), the conditional probabilities simplify to
$P_{\tilde{\bf m}|{\bf m}} = \delta_{\tilde{\bf m}|{\bf m}}$ and $P_{\tilde{\bf m}|{\bf m}} = 1/2^M$, respectively, giving
$\rho_{\tilde{\mathbf{m}}}=
\ket{\psi_{\tilde{\mathbf{m}}}}\bra{\psi_{\tilde{\mathbf{m}}}}$
and
$\rho_{\tilde{\mathbf{m}}} = 
\sum_{\mathbf{m}} P_{\mathbf{m}} \ket{\psi_{\mathbf{m}}}\bra{\psi_{\mathbf{m}}}$,
for the cases of no and maximal assignment errors, respectively. We will refer to density matrix in the presence of maximal noise as $\rho_\varnothing$ since it is like having no measurement record at all or without postselection.

\begin{figure*}
    \centering
    \includegraphics[width=0.99\linewidth]{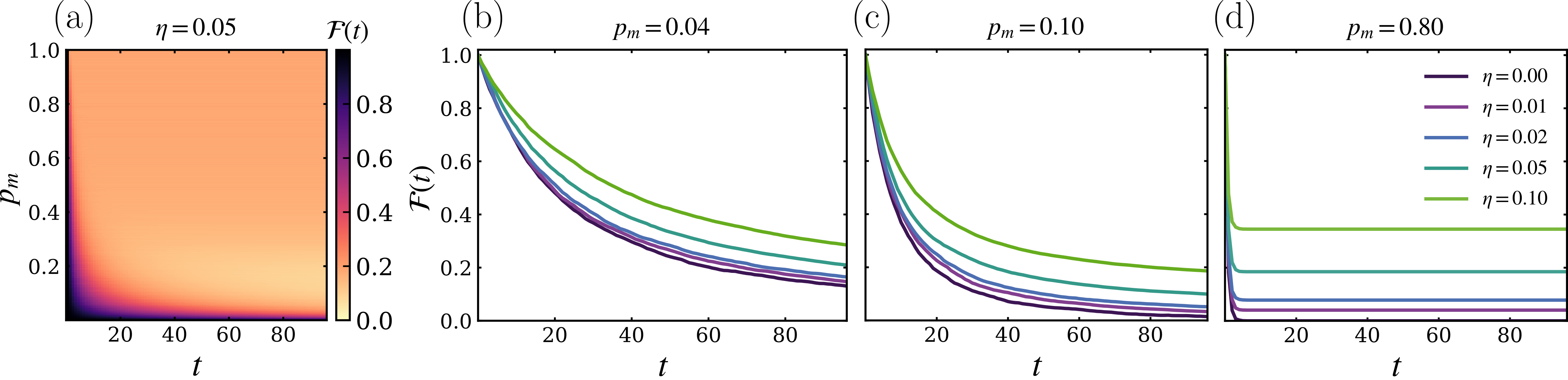}
    \caption{
    Charge-fluctuation dynamics in the presence of readout error.
    (a) Heat map of the normalized charge fluctuation
    ${\cal F}(t)=\overline{\delta^2 Q}(t)/(L/4)$
    as a function of time $t$ and measurement rate $p_m$, at fixed assignment error rate $\eta=5\%$.
    (b),(c),(d) Time evolution of $\mathcal F(t)$ for $p_m=0.04,0.1, 0.8$, respectively.
    Ideal measurements suppress charge fluctuations efficiently, while finite assignment error leaves a residual long-time fluctuation.
    Nonetheless, the qualitative distinction between the two monitoring regimes persists in the presence of noise: $\mathcal F(t)$ decays slowly at low $p_m$ (b) and around the charge sharpening transition $p_m\sim 0.1$ (c), while it saturates rapidly to a finite plateau at high $p_m$ (d).
    All the results are obtained for length $L=12$.
    }
    \label{fig:2}
\end{figure*}

\emph{\color{blue} Charge dynamics under readout noise}.---To study the charge dynamics, we numerically simulate the circuit shown in Fig.~\ref{fig:1}, initialized in an equal superposition of all charge configurations, $\ket{\Psi_0}=\bigotimes_{i=1}^{L}\ket{+}_i$ with $\ket{+}=(\ket{0}+\ket{1})/\sqrt{2}$. The simulation method is briefly described in the End Matter. We quantify the charge dynamics through the mean charge fluctuation
\begin{align}
    \overline{\delta^2 Q} = \mathbb{E}_{\mathcal{U},\mathcal{M},\tilde{\mathbf m}}\left[\delta^2 Q\big|_{\rho_{\tilde{\mathbf m}}}\right],
\end{align}
obtained by averaging $\delta^2 Q|_{\rho_{\tilde{\mathbf m}}}$ over random unitaries ${\cal U}$, measurement locations ${\cal M}$, and readout records $\tilde{\mathbf m}$. Figure~\ref{fig:2} shows the normalized charge fluctuation $\mathcal F(t)=\overline{\delta^2 Q}(t)/(L/4)$ for a moderate error rate $\eta=5\%$. In the ideal case ($\eta=0$), $\mathcal F(t)$ remains finite at long times for $p_m<p_\#$, whereas above the charge-sharpening transition at $p_\#$ it decays monotonically on timescales $t\gtrsim L$ and is eventually fully suppressed.

At a finite assignment-error rate, however, the fluctuations never vanish: $\mathcal F(t)$ instead saturates to a finite plateau (Fig.~\ref{fig:2}), even for $p_m>p_\#$. These residual fluctuations arise because the reported record $\tilde{\mathbf m}$ is compatible with many distinct true histories $\mathbf m$. Postselecting on $\tilde{\mathbf m}$ therefore no longer projects the state onto a single charge sector. The plateau height increases with $\eta$, reflecting how strongly readout noise degrades charge sharpness.
Notably, the error-free histories ($\mathbf m=\tilde{\mathbf m}$) carry only a small fraction of the ensemble weight, $r_{\rm corr}\simeq(1-\eta)^{M}$, where $M\simeq p_m L T$ is the total number of measurements. For the parameters of Fig.~\ref{fig:2} with $\eta=0.05$ and a high measurement rate, $r_{\rm corr}\sim 0.007$, yet $\mathcal F$ remains far below its maximal value $\mathcal F=1$. This is because the charge fluctuations of the mixture are set not by how many wrong trajectories are present in the noisy ensemble but by how widely their charges are spread. Each history is itself nearly charge-sharp, and typical erroneous trajectories differ from $\tilde{\mathbf m}$ in only a few outcomes, so they occupy charge sectors close to that of the error-free one. 

The measurement rate controls both the relaxation toward this noise-induced steady state and its long-time magnitude. From very weak monitoring up to the vicinity of the ideal sharpening transition ($p_m\sim 0.1$) [Figs.~\ref{fig:2}(b),(c)], $\mathcal F(t)$ relaxes slowly; a finite noise-induced plateau is ultimately reached, but it is barely discernible over the long times shown in the figure ($t\gg L$). In contrast, at strong monitoring [Fig.~\ref{fig:2}(d)] the relaxation is much faster, with the steady-state value reached within $t^\ast\sim L$. Importantly, this sharp separation in relaxation timescales persists even in the presence of readout noise: noise prevents the fluctuations from vanishing and produces a finite asymptotic plateau, but it does not erase the qualitative difference between slow relaxation at weak monitoring and rapid, $t^\ast\lesssim L$ relaxation at strong monitoring.

A distinctive feature of readout errors in monitored dynamics is that their cumulative effect depends not only on the error rate and circuit depth, but also on how frequently mid-circuit measurements occur. Less frequent measurements give the trajectory more opportunity to accumulate charge information from correctly reported outcomes before an assignment error corrupts the record; consequently, for the same error rate $\eta$, less frequent measurements yield a smaller additional admixture of charge sectors. At small measurement rates, in the charge-fuzzy phase, we therefore expect the qualitative behavior of the fluctuations to persist under readout noise, since they remain dominated by those inherited from the initial state's broad spread across charge sectors, which readout noise can only slightly enhance. At higher measurement rates, however, the noise-induced partial postselection becomes essentially the sole source of fluctuations, since without noise the charge distribution would be perfectly sharp.

\begin{figure*}
    \centering
    \includegraphics[width=0.99\linewidth]{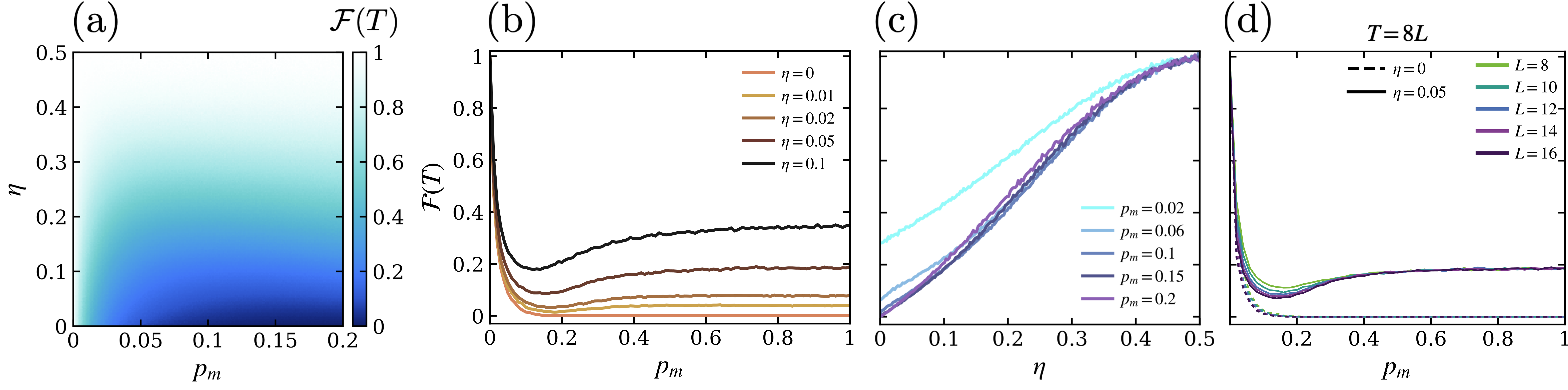}
    \caption{Steady-state charge fluctuations in the presence of readout error. (a) Heat map of the normalized charge fluctuation ${\cal F}(T)$ at long time $T=8L$, as a function of measurement rate $p_m$ (near the sharpening transition, $p_m\sim 0.1$) and error rate $\eta$. (b) Measurement-rate dependence of the fluctuations for various error strengths. (c) Growth of the fluctuations with error strength for various values of $p_m$. Panels (a)--(c) are all for $L=12$; panel (d) shows that other system sizes exhibit the same pattern, with only a slight decrease in the depth of the minimum with increasing size.}   \label{fig:3}
\end{figure*}

These features are summarized in Fig.~\ref{fig:3}, which shows the long-time steady-state fluctuations as a function of both measurement and error rates. Panels (a) and (b) show that for weak-to-moderate noise, the fluctuations closely track the noiseless behavior for $p_m\lesssim p_\#\sim 0.1$, pass through a shallow minimum near the ideal transition point $p_\#$, and then plateau, becoming essentially independent of $p_m$ for $p_m\gtrsim 0.4$. 
The variation of fluctuations with error rate for various fixed $p_m$ is shown in panel (c). Panel (d) demonstrate that system size has only a mild effect on the depth of this minimum at moderate noise.

The shallow dip near $p_\#$ signals a competition between measurement-induced purification and noise-induced mixing, and we argue below that it survives as a footprint of the charge-sharpening transition even in the presence of readout noise. Up to moderate error rates ($\eta\lesssim 0.1$), the dip deepens slightly with increasing $\eta$, since larger errors raise the residual plateau it sits against. This structure (a shallow minimum near $p_\#$ followed by a wide plateau) identifies three distinct regimes as a function of measurement rate: (i) at low $p_m$, fluctuations remain dominated by the initial charge-fuzzy state, and partial postselection under noise is still effective enough to reduce them down to a shallow minimum at $p_m^\ast\sim p_\#$; (ii) at intermediate rates, $p_m^\ast \lesssim p_m \lesssim 0.4$, purification and noise-induced mixing compete directly, with mixing increasingly dominant as $p_m$ grows, so the fluctuations rise; and (iii) at high rates, $p_m\gtrsim 0.4$, the two effects balance, producing a plateau of residual fluctuations independent of $p_m$. This three-regime structure holds until the noise becomes very strong ($\eta\to1/2$), at which point noise-induced mixing suppresses any reduction in the fluctuations entirely and $\mathcal F=1$ for all $p_m$.

\emph{\color{blue} Noise-induced fluctuations}.---Aforementioned competition between noise-induced mixing and the suppression of fluctuations by (partial) postselection can be made precise by decomposing the charge fluctuations into quantum and classical contributions, as detailed in the End Matter. The quantum contribution captures the intra-trajectory fluctuations arising from the quantum uncertainty of charge within each pure state (trajectory) in the ensemble; it coincides exactly with the ideal, noise-free postselected fluctuations, averaged over trajectories and circuit realizations, ${\cal F}_{\rm qu}\equiv {\cal F}_{\eta=0}$, and therefore vanishes deep in the charge-sharp phase. The classical contribution instead captures the inter-trajectory variance -- how much $\langle Q \rangle_{\bf m}$ differs from one record ${\bf m}$ to another within the noisy ensemble -- and is precisely a quantitative measure of what we have so far called noise-induced mixing. This separation clarifies the origin of regime (ii) above: once the quantum fluctuations are fully suppressed, the modest growth of the total fluctuations with $p_m$ reflects only the growing classical contribution. This regime must therefore set in just past the sharpening transition, which is why, even in our finite-size simulations, the minimum persists near the true transition point, $p_m^\ast\sim p_\#$.

\begin{figure}
    \centering
    \includegraphics[width=0.99\linewidth]{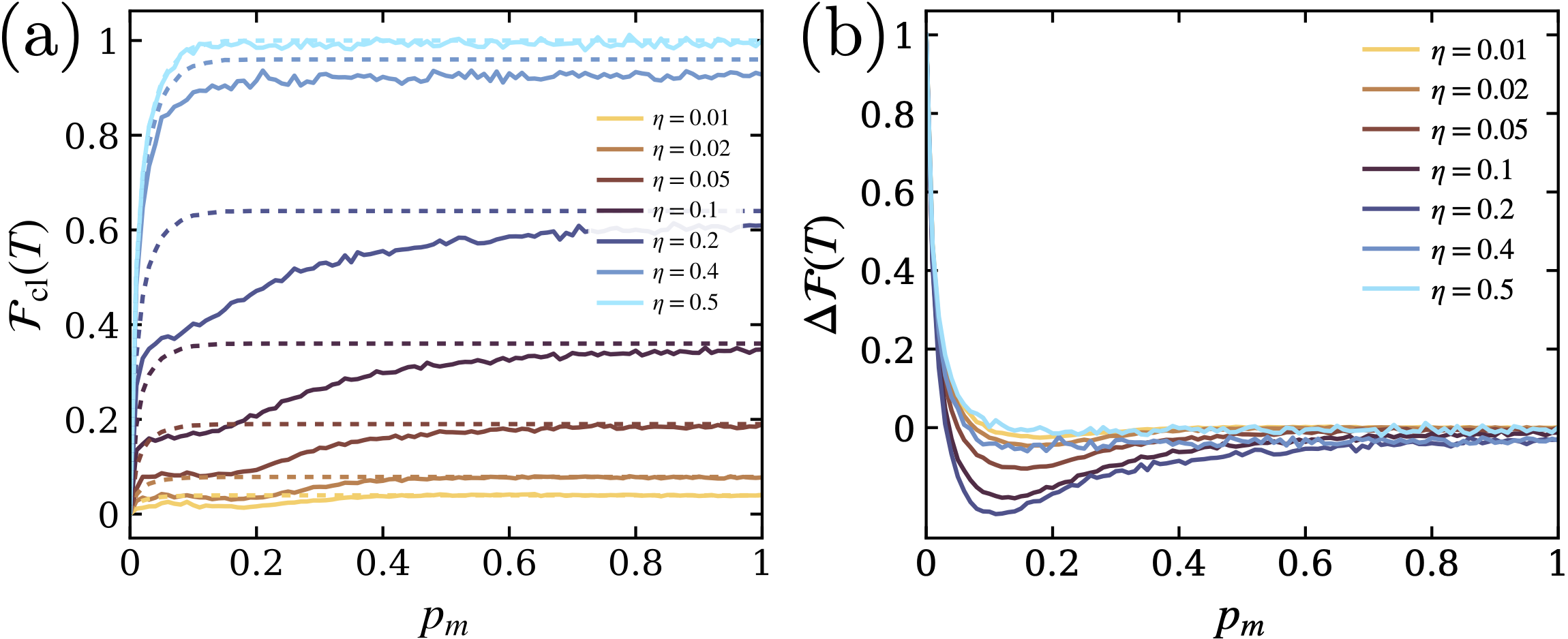} 
    \caption{ Noise-induced mixing and its associated extra fluctuations. (a) The noise-induced (classical) contribution ${\cal F}_{\rm cl}$ as a function of measurement rate $p_m$, for different error rates (solid lines); the reason for calling this term classical is explained in the main text. Dashed lines show the corresponding prediction assuming uncorrelated errors, governed by the Bernoulli noise-variance form (denoted ${\cal F}^{\rm approx}_{\rm cl}$ in the main text). (b) The residual fluctuation after subtracting the uncorrelated classical estimate from the total, $\Delta{\cal F} = {\cal F} - {\cal F}^{\rm approx}_{\rm cl}$. All the results are obtained for length $L=12$.
}
    \label{fig:4}
\end{figure}

To see how this separation accounts for the overall trend, Fig.~\ref{fig:4}(a) shows the classical contribution, ${\cal F}_{\rm cl} \equiv {\cal F} - {\cal F}_{\eta=0}$, as a function of $p_m$. It grows monotonically with measurement rate, as expected since more frequent measurements produce a more strongly mixed ensemble. But the growth rate itself changes across three regimes: rapid growth at small $p_m$, a more moderate rate at intermediate $p_m$, and saturation to a maximal value at large $p_m$. The dashed lines in Fig.~\ref{fig:4}(a) show an approximate form,
\begin{align}
{\cal F}^{\rm approx}_{\rm cl}(p_m,\eta) = {\cal E}(\eta)\,{\cal F}_{\rm cl}(p_m,\eta=1/2),
\end{align}
using the Bernoulli noise-variance factor ${\cal E}(\eta)=4\eta(1-\eta)$ appropriate for completely uncorrelated errors. This approximation works well at large $p_m$, where the classical contribution saturates, but deviates in the intermediate regime -- precisely where the total fluctuations grow past the minimum. The assignment errors themselves are independent and uncorrelated by construction in our model; however, the trajectory weights $P_{\bf m}$ in Eq.~\eqref{eq:noisy_ensemble} carry memory from the full monitored dynamics, which introduces effective correlations into how errors accumulate within a given noisy ensemble. Our results show that these correlations are strongest at intermediate measurement rates, where they suppress the noise-induced mixing relative to the fully uncorrelated estimate (dashed lines).

At high measurement rates, by contrast, only the most recent outcomes and their associated errors matter, and the noisy ensemble is well described by an effectively uncorrelated noise channel. To quantify the deviation from this picture, we define $\Delta{\cal F} = {\cal F} - {\cal F}^{\rm approx}_{\rm cl}$, the difference between the total noisy fluctuations and the uncorrelated classical estimate; negative values signal an attenuation of noise-induced mixing relative to the uncorrelated approximation. This dip in $\Delta{\cal F}$ should also lie close to the sharpening transition, where trajectory history and its associated correlations are expected to play the strongest role in suppressing the classical fluctuations.

\emph{\color{blue} Discussion}.---Our results demonstrate that finite readout error does not destroy the charge-sharpening transition, but converts the sharp phase's vanishing fluctuations into a finite, $\eta$-dependent plateau while preserving the timescale separation between the two phases. A perhaps more interesting feature is the shallow minimum that appears near $p_\#$ in the long-time fluctuations as a function of measurement rate $p_m$. The quantum/classical decomposition explains this minimum directly. It marks the point where intrinsic quantum fluctuations have already been suppressed by postselection, while noise-induced classical mixing has not yet taken over, a fact that can serve as a fingerprint of the ideal transition surviving inside a noisy signal.

While the classical/quantum split is conceptually clarifying, ${\cal F}_{\rm cl}$ is not obviously easier to measure directly. The uncorrelated approximation nonetheless offers a practical alternative. Since ${\cal E}(\eta)$ is fixed entirely by the known readout fidelity, subtracting ${\cal E}(\eta)\mathcal F_{\rm cl}(p_m,\eta=1/2)$ from the total $\mathcal F$ yields $\Delta\mathcal F$, a hybrid theory-experiment probe whose $p_m$-dependence comes entirely from the underlying noisy dynamics combined with an idealized theoretical input. This offers a route to locate the transition and distinguish fuzzy from sharp behavior in the presence of readout noise.

Finally, our results point to a broader issue. Readout errors on mid-circuit measurements are not equivalent to standard end-of-circuit readout errors, which a confusion matrix applied after the fact can cleanly undo. Here, corrupted outcomes determine which state the trajectory is postselected onto during the dynamics, so the noise contaminates the conditioning itself rather than only the final readout. Mitigating such errors is, to the best of our knowledge, a qualitatively harder and largely open problem, and our results offer a first step in this direction. The uncorrelated-error model ${\cal F}^{\rm approx}_{\rm cl}$ already captures the total noisy fluctuations well at both low and high $p_m$ [Fig.~\ref{fig:4}(b)]. Its failure at intermediate $p_m$, where $\Delta{\cal F}<0$, singles out the regime in which memory retained through partial postselection makes the effective error channel strongly non-Markovian. Capturing these memory effects is thus the central challenge for mitigating mid-circuit readout errors that warrants further exploration in the future.

\emph{\color{blue} Conclusions}.---We have shown that charge sharpening in U(1)-symmetric monitored circuits survives readout errors, with the sharp phase's vanishing fluctuations replaced by a finite plateau
and a shallow minimum near ideal transition point $p_\#$. Applying a quantum/classical decomposition to the fluctuations, the existence of minimum is attributed to a competition between measurement-induced purification and noise-induced mixing, with intermediate-$p_m$ deviations from the uncorrelated-noise form signaling memory retained through  partial (incomplete) postselection. Since mid-circuit errors corrupt the postselection itself, they resist standard end-of-circuit mitigation; our decomposition offers a minimal alternative for locating the transition directly from experimental data. More broadly, these results establish charge fluctuations as a robust probe of measurement-induced criticality under realistic readout noise, and point to noise mitigation for mid-circuit measurements as an open direction for future work.

\emph{\color{blue} Acknowledgments}.---M.A.M. acknowledges the financial support of the Finnish Ministry of Education and Culture through the Quantum Doctoral Education Pilot
Program (QDOC VN/3137/2024-OKM-4) and the Research Council of Finland through the Finnish Quantum Flagship project (TAU 359020). T.O. acknowledges support from the Finnish Research Council project 362573 and the Finnish quantum flagship program. A.G.M. acknowledges the financial support from the Finnish Center of Excellence in Quantum Materials QMAT (No. 374166), and the ERC Consolidator Grant ULTRATWISTROICS (No. 101170477).

\bibliography{refs}

@misc{code_zenodo,
  author       = {Mir, Murtaza Ali and Moghaddam, Ali G.},
  title        = {{Readout errors and incomplete postselection in monitored quantum circuits}},
  howpublished = {Zenodo: \href{https://doi.org/10.5281/zenodo.23372650}{10.5281/zenodo.23372650}},
  year         = {2026}
}

@article{Wei2019imperfect_measurement,
  title = {Detector tomography on IBM quantum computers and mitigation of an imperfect measurement},
  author = {Chen, Yanzhu and Farahzad, Maziar and Yoo, Shinjae and Wei, Tzu-Chieh},
  journal = {Phys. Rev. A},
  volume = {100},
  issue = {5},
  pages = {052315},
  numpages = {17},
  year = {2019},
  month = {Nov},
  publisher = {American Physical Society},
  doi = {10.1103/PhysRevA.100.052315},
  url = {https://link.aps.org/doi/10.1103/PhysRevA.100.052315}
}

@article{Maciejewski2021modelingmitigation,
  doi = {10.22331/q-2021-06-01-464},
  url = {https://doi.org/10.22331/q-2021-06-01-464},
  title = {Modeling and mitigation of cross-talk effects in readout noise with applications to the {Q}uantum {A}pproximate {O}ptimization {A}lgorithm},
  author = {Maciejewski, Filip B. and Baccari, Flavio and Zimbor{\'{a}}s, Zolt{\'{a}}n and Oszmaniec, Micha{\l{}}},
  journal = {{Quantum}},
  issn = {2521-327X},
  publisher = {{Verein zur F{\"{o}}rderung des Open Access Publizierens in den Quantenwissenschaften}},
  volume = {5},
  pages = {464},
  month = jun,
  year = {2021}
}

@article{Agrawal2024Observing_Learnability,
  title = {Observing Quantum Measurement Collapse as a Learnability Phase Transition},
  author = {Agrawal, Utkarsh and Lopez-Piqueres, Javier and Vasseur, Romain and Gopalakrishnan, Sarang and Potter, Andrew C.},
  journal = {Phys. Rev. X},
  volume = {14},
  issue = {4},
  pages = {041012},
  numpages = {12},
  year = {2024},
  month = {Oct},
  publisher = {American Physical Society},
  doi = {10.1103/PhysRevX.14.041012},
  url = {https://link.aps.org/doi/10.1103/PhysRevX.14.041012}
}

@article{Vasseur2022learnability,
  title = {Transitions in the Learnability of Global Charges from Local Measurements},
  author = {Barratt, Fergus and Agrawal, Utkarsh and Potter, Andrew C. and Gopalakrishnan, Sarang and Vasseur, Romain},
  journal = {Phys. Rev. Lett.},
  volume = {129},
  issue = {20},
  pages = {200602},
  numpages = {7},
  year = {2022},
  month = {Nov},
  publisher = {American Physical Society},
  doi = {10.1103/PhysRevLett.129.200602},
  url = {https://link.aps.org/doi/10.1103/PhysRevLett.129.200602}
}

@article{google2023measurement,
  title = {Measurement-induced entanglement and teleportation on a noisy quantum processor},
  author = {Hoke~\textit{et al.,}~{(Google Quantum AI and Collaborators)} , J. C. },
  journal = {Nature},
  volume = {622},
  issue = {7983},
  pages = {481},
  numpages = {6},
  year = {2023},
  month = {Oct},
  publisher = {Springer Nature Limited},
  doi = {10.1038/s41586-023-06505-7},
  url = {https://www.nature.com/articles/s41586-023-06505-7}
}

@article{Ippoliti_postselection,
  title = {Postselection-Free Entanglement Dynamics via Spacetime Duality},
  author = {Ippoliti, Matteo and Khemani, Vedika},
  journal = {Phys. Rev. Lett.},
  volume = {126},
  issue = {6},
  pages = {060501},
  numpages = {7},
  year = {2021},
  month = {Feb},
  publisher = {American Physical Society},
  doi = {10.1103/PhysRevLett.126.060501},
  url = {https://link.aps.org/doi/10.1103/PhysRevLett.126.060501}
}

@article{Fisher_random_quantum_circuits,
author = {Fisher, Matthew P.A. and Khemani, Vedika and Nahum, Adam and Vijay, Sagar},
title = {Random Quantum Circuits},
journal = {Annu. Rev. Condens. Matter Phys.},
volume = {14},
number = {1},
pages = {335},
year = {2023},
doi = {10.1146/annurev-conmatphys-031720-030658}
}

@article{Jian-Ludwig2020_measurement_criticality,
  title = {Measurement-induced criticality in random quantum circuits},
  author = {Jian, Chao-Ming and You, Yi-Zhuang and Vasseur, Romain and Ludwig, Andreas W. W.},
  journal = {Phys. Rev. B},
  volume = {101},
  issue = {10},
  pages = {104302},
  numpages = {11},
  year = {2020},
  month = {Mar},
  publisher = {American Physical Society},
  doi = {10.1103/PhysRevB.101.104302},
  url = {https://link.aps.org/doi/10.1103/PhysRevB.101.104302}
}

@article{Pal_2020_MBL_MIPT,
  title = {Measurement-induced entanglement transitions in many-body localized systems},
  author = {Lunt, Oliver and Pal, Arijeet},
  journal = {Phys. Rev. Res.},
  volume = {2},
  issue = {4},
  pages = {043072},
  numpages = {10},
  year = {2020},
  month = {Oct},
  publisher = {American Physical Society},
  doi = {10.1103/PhysRevResearch.2.043072},
  url = {https://link.aps.org/doi/10.1103/PhysRevResearch.2.043072}
}

@article{skinner_measurement-induced_2019,
	title = {Measurement-{Induced} {Phase} {Transitions} in the {Dynamics} of {Entanglement}},
	volume = {9},
	url = {https://link.aps.org/doi/10.1103/PhysRevX.9.031009},
	doi = {10.1103/PhysRevX.9.031009},
	number = {3},
	urldate = {2022-04-22},
	journal = {Phys. Rev. X},
	author = {Skinner, Brian and Ruhman, Jonathan and Nahum, Adam},
	month = jul,
	year = {2019},
	pages = {031009},
}

@article{nahum_measurement_2021,
	title = {Measurement and {Entanglement} {Phase} {Transitions} in {All}-{To}-{All} {Quantum} {Circuits}, on {Quantum} {Trees}, and in {Landau}-{Ginsburg} {Theory}},
	volume = {2},
	url = {https://link.aps.org/doi/10.1103/PRXQuantum.2.010352},
	doi = {10.1103/PRXQuantum.2.010352},
	number = {1},
	urldate = {2022-04-22},
	journal = {PRX Quantum},
	author = {Nahum, Adam and Roy, Sthitadhi and Skinner, Brian and Ruhman, Jonathan},
	month = mar,
	year = {2021},
	pages = {010352},
}

@article{Chakraborty2024,
  title = {{Charge and entanglement criticality in a U(1)-symmetric hybrid circuit of qubits}},
  author = {Chakraborty, Ahana and Chen, Kun and Zabalo, Aidan and Wilson, Justin H. and Pixley, J. H.},
  journal = {Phys. Rev. B},
  volume = {110},
  issue = {4},
  pages = {045135},
  numpages = {17},
  year = {2024},
  month = {Jul},
  publisher = {American Physical Society},
  doi = {10.1103/PhysRevB.110.045135},
  url = {https://link.aps.org/doi/10.1103/PhysRevB.110.045135}
}

@article{li_measurement-driven_2019,
	title = {Measurement-driven entanglement transition in hybrid quantum circuits},
	volume = {100},
	url = {https://link.aps.org/doi/10.1103/PhysRevB.100.134306},
	doi = {10.1103/PhysRevB.100.134306},
	number = {13},
	urldate = {2022-04-22},
	journal = {Phys. Rev. B},
	author = {Li, Yaodong and Chen, Xiao and Fisher, Matthew P. A.},
	month = oct,
	year = {2019},
	pages = {134306},
}

@article{li_quantum_2018,
	title = {Quantum {Zeno} effect and the many-body entanglement transition},
	volume = {98},
	url = {https://link.aps.org/doi/10.1103/PhysRevB.98.205136},
	doi = {10.1103/PhysRevB.98.205136},
	number = {20},
	urldate = {2022-04-22},
	journal = {Phys. Rev. B},
	author = {Li, Yaodong and Chen, Xiao and Fisher, Matthew P. A.},
	month = nov,
	year = {2018},
	pages = {205136},
}

@article{ippoliti_entanglement_2021,
	title = {Entanglement {Phase} {Transitions} in {Measurement}-{Only} {Dynamics}},
	volume = {11},
	url = {https://link.aps.org/doi/10.1103/PhysRevX.11.011030},
	doi = {10.1103/PhysRevX.11.011030},
	number = {1},
	urldate = {2022-04-22},
	journal = {Phys. Rev. X},
	author = {Ippoliti, Matteo and Gullans, Michael J. and Gopalakrishnan, Sarang and Huse, David A. and Khemani, Vedika},
	month = feb,
	year = {2021},
	pages = {011030},
}

@article{bao_theory_2020,
	title = {Theory of the phase transition in random unitary circuits with measurements},
	volume = {101},
	url = {https://link.aps.org/doi/10.1103/PhysRevB.101.104301},
	doi = {10.1103/PhysRevB.101.104301},
	number = {10},
	urldate = {2022-04-22},
	journal = {Phys. Rev. B},
	author = {Bao, Yimu and Choi, Soonwon and Altman, Ehud},
	month = mar,
	year = {2020},
	pages = {104301},
}

@article{choi_quantum_2020,
	title = {Quantum {Error} {Correction} in {Scrambling} {Dynamics} and {Measurement}-{Induced} {Phase} {Transition}},
	volume = {125},
	url = {https://link.aps.org/doi/10.1103/PhysRevLett.125.030505},
	doi = {10.1103/PhysRevLett.125.030505},
	number = {3},
	urldate = {2022-04-22},
	journal = {Phys. Rev. Lett.},
	author = {Choi, Soonwon and Bao, Yimu and Qi, Xiao-Liang and Altman, Ehud},
	month = jul,
	year = {2020},
	pages = {030505},
}

@article{gullans_dynamical_2020,
	title = {Dynamical {Purification} {Phase} {Transition} {Induced} by {Quantum} {Measurements}},
	volume = {10},
	url = {https://link.aps.org/doi/10.1103/PhysRevX.10.041020},
	doi = {10.1103/PhysRevX.10.041020},
	number = {4},
	urldate = {2022-04-22},
	journal = {Phys. Rev. X},
	author = {Gullans, Michael J. and Huse, David A.},
	month = oct,
	year = {2020},
	pages = {041020},
}

@article{Turkeshi2022_data,
  title = {Measurement-induced criticality as a data-structure transition},
  author = {Turkeshi, Xhek},
  journal = {Phys. Rev. B},
  volume = {106},
  issue = {14},
  pages = {144313},
  numpages = {9},
  year = {2022},
  month = {Oct},
  publisher = {American Physical Society},
  doi = {10.1103/PhysRevB.106.144313},
  url = {https://link.aps.org/doi/10.1103/PhysRevB.106.144313}
}

@article{Marino2025_symmetry,
  title = {Generalizing measurement-induced phase transitions to information exchange symmetry breaking},
  author = {Kelly, Shane P. and Marino, Jamir},
  journal = {Phys. Rev. A},
  volume = {111},
  issue = {1},
  pages = {012425},
  numpages = {26},
  year = {2025},
  month = {Jan},
  publisher = {American Physical Society},
  doi = {10.1103/PhysRevA.111.012425},
  url = {https://link.aps.org/doi/10.1103/PhysRevA.111.012425}
}

@article{Agrawal_2022,
	url = {https://doi.org/10.1103\%2Fphysrevx.12.041002},
	year = 2022,
	month = {oct},
	publisher = {American Physical Society ({APS})},
	volume = {12},
	number = {4},
	author = {Utkarsh Agrawal and Aidan Zabalo and Kun Chen and Justin H. Wilson and Andrew C. Potter and J.H. Pixley and Sarang Gopalakrishnan and Romain Vasseur},
	title = {{Entanglement and Charge-Sharpening Transitions in U(1) Symmetric Monitored Quantum Circuits}},
	journal = {Phys. Rev. X}
}

@inbook{potter_entanglement_2021,
	address = {Cham},
	author = {Potter, Andrew C. and Vasseur, Romain},
	booktitle = {Entanglement in Spin Chains: From Theory to Quantum Technology Applications},
	doi = {10.1007/978-3-031-03998-0_9},
	editor = {Bayat, Abolfazl and Bose, Sougato and Johannesson, Henrik},
	isbn = {978-3-031-03998-0},
	pages = {211--249},
	publisher = {Springer International Publishing},
	title = {Entanglement Dynamics in Hybrid Quantum Circuits},
	url = {https://doi.org/10.1007/978-3-031-03998-0_9},
	year = {2022},
}

@article{poyhonen2024,
      title={Scalable approach to monitored quantum dynamics and entanglement phase transitions}, 
      author={P\"oyh\"onen, Kim and Moghaddam, Ali G. and Ivaki, Moein N. and Ojanen, Teemu},
      year={2024},
      journal = {arXiv:2406.19052},
      url={https://arxiv.org/abs/2406.19052}
}

@article{XEB-altman,
  title = {Cross Entropy Benchmark for Measurement-Induced Phase Transitions},
  author = {Li, Yaodong and Zou, Yijian and Glorioso, Paolo and Altman, Ehud and Fisher, Matthew P. A.},
  journal = {Phys. Rev. Lett.},
  volume = {130},
  issue = {22},
  pages = {220404},
  numpages = {6},
  year = {2023},
  month = {Jun},
  publisher = {American Physical Society},
  doi = {10.1103/PhysRevLett.130.220404},
  url = {https://link.aps.org/doi/10.1103/PhysRevLett.130.220404}
}

@misc{XEB-experiment,
      title={Experimental demonstration of scalable cross-entropy benchmarking to detect measurement-induced phase transitions on a superconducting quantum processor}, 
      author={Hirsh Kamakari and Jiace Sun and Yaodong Li and Jonathan J. Thio and Tanvi P. Gujarati and Matthew P. A. Fisher and Mario Motta and Austin J. Minnich},
      year={2024},
      eprint={2403.00938},
      archivePrefix={arXiv},
      primaryClass={quant-ph}
}

@article{Minnich_2023measurement,
  title={Measurement-induced entanglement phase transition on a superconducting quantum processor with mid-circuit readout},
  author={Koh, Jin Ming and Sun, Shi-Ning and Motta, Mario and Minnich, Austin J},
  journal={Nat. Phys.},
  volume={19},
  number={9},
  pages={1314},
  year={2023},
  doi={10.1038/s41567-023-02076-6},
  publisher={Nature Publishing Group UK London}
}

@article{Monroe_measurement_2022,
	doi = {10.1038/s41567-022-01619-7},
	year = 2022,
	month = {jun},
	publisher = {Springer Science and Business Media {LLC}},
	volume = {18},
	number = {7},
	pages = {760--764},
	author = {Crystal Noel and Pradeep Niroula and Daiwei Zhu and Andrew Risinger and Laird Egan and Debopriyo Biswas and Marko Cetina and Alexey V. Gorshkov and Michael J. Gullans and David A. Huse and Christopher Monroe},
	title = {Measurement-induced quantum phases realized in a trapped-ion quantum computer},
	journal = {Nat. Phys.}
}

@article{Halpern2023,
  title = {Critical phase and spin sharpening in SU(2)-symmetric monitored quantum circuits},
  author = {Majidy, Shayan and Agrawal, Utkarsh and Gopalakrishnan, Sarang and Potter, Andrew C. and Vasseur, Romain and Halpern, Nicole Yunger},
  journal = {Phys. Rev. B},
  volume = {108},
  issue = {5},
  pages = {054307},
  numpages = {13},
  year = {2023},
  month = {Aug},
  publisher = {American Physical Society},
  doi = {10.1103/PhysRevB.108.054307},
  url = {https://link.aps.org/doi/10.1103/PhysRevB.108.054307}
}

@article{Michailidis20revivals,
  title = {Slow Quantum Thermalization and Many-Body Revivals from Mixed Phase Space},
  author = {Michailidis, A. A. and Turner, C. J. and Papi\ifmmode \acute{c}\else \'{c}\fi{}, Z. and Abanin, D. A. and Serbyn, M.},
  journal = {Phys. Rev. X},
  volume = {10},
  issue = {1},
  pages = {011055},
  numpages = {28},
  year = {2020},
  month = {Mar},
  publisher = {American Physical Society},
  doi = {10.1103/PhysRevX.10.011055},
  url = {https://link.aps.org/doi/10.1103/PhysRevX.10.011055}
}

@article{ippoliti2023learnability,
  title = {Learnability Transitions in Monitored Quantum Dynamics via Eavesdropper's Classical Shadows},
  author = {Ippoliti, Matteo and Khemani, Vedika},
  journal = {PRX Quantum},
  volume = {5},
  issue = {2},
  pages = {020304},
  numpages = {24},
  year = {2024},
  month = {Apr},
  publisher = {American Physical Society},
  doi = {10.1103/PRXQuantum.5.020304},
  url = {https://link.aps.org/doi/10.1103/PRXQuantum.5.020304}
}

@article{PRXQuantum.4.040309,
  title = {Measurement-Induced Phases of Matter Require Feedback},
  author = {Friedman, Aaron J. and Hart, Oliver and Nandkishore, Rahul},
  journal = {PRX Quantum},
  volume = {4},
  issue = {4},
  pages = {040309},
  numpages = {46},
  year = {2023},
  month = {Oct},
  publisher = {American Physical Society},
  doi = {10.1103/PRXQuantum.4.040309},
  url = {https://link.aps.org/doi/10.1103/PRXQuantum.4.040309}
}

@article{Gullans2024magic,
  title={Phase transition in magic with random quantum circuits},
  author={Niroula, Pradeep and White, Christopher David and Wang, Qingfeng and Johri, Sonika and Zhu, Daiwei and Monroe, Christopher and Noel, Crystal and Gullans, Michael J},
  journal={Nature physics},
  volume={20},
  number={11},
  pages={1786--1792},
  year={2024},
  doi = {10.1038/s41567-024-02637-3},
  publisher={Nature Publishing Group UK London}
}

@article{Yamamoto2023,
  title = {Localization properties in disordered quantum many-body dynamics under continuous measurement},
  author = {Yamamoto, Kazuki and Hamazaki, Ryusuke},
  journal = {Phys. Rev. B},
  volume = {107},
  issue = {22},
  pages = {L220201},
  numpages = {7},
  year = {2023},
  month = {Jun},
  publisher = {American Physical Society},
  doi = {10.1103/PhysRevB.107.L220201},
  url = {https://link.aps.org/doi/10.1103/PhysRevB.107.L220201}
}

@article{Fazio2024postselection,
  title = {Many-Body Dynamics in Monitored Atomic Gases without Postselection Barrier},
  author = {Passarelli, Gianluca and Turkeshi, Xhek and Russomanno, Angelo and Lucignano, Procolo and Schir\`o, Marco and Fazio, Rosario},
  journal = {Phys. Rev. Lett.},
  volume = {132},
  issue = {16},
  pages = {163401},
  numpages = {7},
  year = {2024},
  month = {Apr},
  publisher = {American Physical Society},
  doi = {10.1103/PhysRevLett.132.163401},
  url = {https://link.aps.org/doi/10.1103/PhysRevLett.132.163401}
}

@article{Mirlin2023,
  title = {Theory of Free Fermions under Random Projective Measurements},
  author = {Poboiko, Igor and P\"opperl, Paul and Gornyi, Igor V. and Mirlin, Alexander D.},
  journal = {Phys. Rev. X},
  volume = {13},
  issue = {4},
  pages = {041046},
  numpages = {26},
  year = {2023},
  month = {Dec},
  publisher = {American Physical Society},
  doi = {10.1103/PhysRevX.13.041046},
  url = {https://link.aps.org/doi/10.1103/PhysRevX.13.041046}
}

@article{Zarand2025,
  title = {{Nonstabilizerness in open XXZ spin chains: Universal scaling and dynamics}},
  author ={Sticlet, Doru and D{\'o}ra, Bal{\'a}zs and Szombathy, Dominik and Zar{\'a}nd, Gergely and Moca, C{\u{a}}t{\u{a}}lin Pa{\c{s}}cu},
  journal = {Phys. Rev. Res.},
  volume = {7},
  issue = {4},
  pages = {043130},
  numpages = {7},
  year = {2025},
  month = {Nov},
  publisher = {American Physical Society},
  doi = {10.1103/96bk-xf8p},
  url = {https://link.aps.org/doi/10.1103/96bk-xf8p}
}

@article{McGinley2024,
  title = {Postselection-Free Learning of Measurement-Induced Quantum Dynamics},
  author = {McGinley, Max},
  journal = {PRX Quantum},
  volume = {5},
  issue = {2},
  pages = {020347},
  numpages = {31},
  year = {2024},
  month = {May},
  publisher = {American Physical Society},
  doi = {10.1103/PRXQuantum.5.020347},
  url = {https://link.aps.org/doi/10.1103/PRXQuantum.5.020347}
}

@article{vasseur2026houches,
  title={Les Houches lectures on random quantum circuits and monitored quantum dynamics},
  author={Vasseur, Romain},
  journal={arXiv:2602.17258},
  doi = {10.48550/arXiv.2602.17258},
  year={2026}
}

@article{Gambetta2021ScalableM3,
  title     = {Scalable mitigation of measurement errors on quantum computers},
  author    = {Nation, Paul D. and Kang, Hwajung and Sundaresan, Neereja and Gambetta, Jay M.},
  journal   = {Physical Review A},
  volume    = {104},
  number    = {5},
  pages     = {052420},
  year      = {2021},
  month     = {nov},
  publisher = {American Physical Society},
  doi       = {10.1103/PhysRevA.104.052420},
  url       = {https://arxiv.org/abs/2108.12518},
  note      = {arXiv preprint arXiv:2108.12518}
}

@article{mitigation_RMP,
  title = {Quantum error mitigation},
  author = {Cai, Zhenyu and Babbush, Ryan and Benjamin, Simon C. and Endo, Suguru and Huggins, William J. and Li, Ying and McClean, Jarrod R. and O'Brien, Thomas E.},
  journal = {Rev. Mod. Phys.},
  volume = {95},
  issue = {4},
  pages = {045005},
  numpages = {37},
  year = {2023},
  month = {Dec},
  publisher = {American Physical Society},
  doi = {10.1103/RevModPhys.95.045005},
  url = {https://link.aps.org/doi/10.1103/RevModPhys.95.045005}
}

@article{Nachman2020Unfolding,
  title     = {Unfolding quantum computer readout noise},
  author    = {Nachman, Benjamin and Urb{\'a}nek, Miroslav and de Jong, Wibe A. and Bauer, Christian W.},
  journal   = {npj Quantum Information},
  volume    = {6},
  number    = {1},
  pages     = {84},
  year      = {2020},
  month     = {sep},
  publisher = {Nature Publishing Group},
  doi       = {10.1038/s41534-020-00309-7},
  url       = {https://www.nature.com/articles/s41534-020-00309-7},
  note      = {arXiv preprint arXiv:1910.01969}
}

@article{Nachman_activereadout,
  title = {Active readout-error mitigation},
  author = {Hicks, Rebecca and Kobrin, Bryce and Bauer, Christian W. and Nachman, Benjamin},
  journal = {Phys. Rev. A},
  volume = {105},
  issue = {1},
  pages = {012419},
  numpages = {13},
  year = {2022},
  month = {Jan},
  publisher = {American Physical Society},
  doi = {10.1103/PhysRevA.105.012419},
  url = {https://link.aps.org/doi/10.1103/PhysRevA.105.012419}
}

@article{Kawabata2026_symmetry,
  title = {Symmetry and topology of monitored quantum dynamics},
  author = {Xiao, Zhenyu and Kawabata, Kohei},
  journal = {Phys. Rev. B},
  volume = {113},
  issue = {13},
  pages = {134307},
  numpages = {16},
  year = {2026},
  month = {Apr},
  publisher = {American Physical Society},
  doi = {10.1103/1q83-jcbw},
  url = {https://link.aps.org/doi/10.1103/1q83-jcbw}
}

\vspace{1cm}

\section{End Matter}

\emph{\color{blue} Simulation method}.---We simulate the monitored dynamics exactly within a pure-state (quantum-trajectory) framework. Each circuit realization is specified by a set of random gates $\mathcal U$ and measurement locations $\mathcal M$, with each site measured with probability $p_m$. In the ideal case, measuring a site yields outcome $m\in\{0,1\}$ with Born probability $\pi_m=\bra{\Psi}\Pi_m\ket{\Psi}$, where $\Pi_m=\ket{m}\bra{m}$, after which the state is projected and renormalized, $\ket{\Psi}\to\Pi_m\ket{\Psi}/\sqrt{\pi_m}$. Each outcome sequence $\mathbf m=(m_1,\dots,m_M)$ defines a quantum trajectory.

Assignment errors cause the apparatus to report $m'$ when the true outcome is $m$, with probability $p_{m'|m}=\Gamma_{m'm}$ given by Eq.~\eqref{eq:stochastic-assignemnt}. Conditioned on a reported outcome $m'$, the state is updated as
\begin{align}
    \rho \to \frac{\sum_{m} p_{m'|m}\,\Pi_m\rho\,\Pi_m}{\sum_{m} p_{m'|m}\,\pi_m}.
\end{align}
We unravel this update stochastically: for a reported outcome $m'$, we apply $\Pi_m$ with the posterior probability
\begin{align}
\frac{p_{m'|m}\,\pi_m}{\sum_{m=0,1} p_{m'|m}\,\pi_m},
\end{align}
so that the true outcome is either $m'$ or its complement $\bar m'$. For fixed $\mathcal U$, $\mathcal M$, and reported record $\tilde{\mathbf m}$, we repeat this procedure $N_{\rm err}$ times, obtaining states $\ket{\Psi_i}$, and estimate $\rho_{\tilde{\mathbf m}} \simeq N_{\rm err}^{-1}\sum_{i}\ket{\Psi_i}\bra{\Psi_i}$, from which we compute $\delta^2 Q|_{\rho_{\tilde{\mathbf m}}}$. Averaging over $\mathcal U$, $\mathcal M$, and $\tilde{\mathbf m}$ then yields $\overline{\delta^2 Q}$.
The simulation code used to obtain the results in this work is openly available on Zenodo \cite{code_zenodo}.

\emph{\color{blue} Separating quantum and classical fluctuations}.---Here, we derive the decomposition of the noisy charge variance into quantum and classical contributions used in the main text, and analyze its behavior in the maximal-readout-noise limit. We use simplified notation for the ideal, postselected expectation value of an observable $A$,
$\langle A \rangle_{\bf m} = \bra{\psi_{\bf m}} A \ket{\psi_{\bf m}}$,
and for the noisy ensemble density matrix,
$\rho_{\tilde{\bf m}} = \sum_{\bf m} c^{\tilde{\bf m}}_{\bf m} \ket{\psi_{\bf m}}\bra{\psi_{\bf m}}$, with weights
\begin{align}
    c^{\tilde{\bf m}}_{\bf m} = \frac{P_{\tilde{\bf m}|{\bf m}} P_{\bf m}}{\sum_{\bf m} P_{\tilde{\bf m}|{\bf m}} P_{\bf m}}
\end{align}
as in Eq.~\eqref{eq:noisy_ensemble}. Expanding the variance of an observable $Q$ in the postselected noisy ensemble $\rho_{\tilde{\bf m}}$,
\begin{align}
    \delta^2 Q|_{\rho_{\tilde{\bf m}}} &= {\rm Tr}\left(Q^2 \rho_{\tilde{\bf m}}\right) - \left[{\rm Tr}\left( Q\rho_{\tilde{\bf m}}\right)\right]^2 \nonumber\\
    &= \sum_{\bf m} c^{\tilde{\bf m}}_{\bf m}\, \langle Q^2 \rangle_{\bf m}
    - \left(\sum_{\bf m} c^{\tilde{\bf m}}_{\bf m}\, \langle Q \rangle_{\bf m} \right)^2,
\end{align}
separates the ideal quantum charge fluctuations from the additional broadening induced by readout errors:
\begin{align}\label{eq:separation}
    \overline{\delta^2 Q} = \overline{{\delta^2 Q}}_{\rm qu} + \overline{{\delta^2 Q}}_{\rm cl}.
\end{align}
This follows by adding and subtracting $\sum_{\bf m} c^{\tilde{\bf m}}_{\bf m}\, \langle Q \rangle_{\bf m}^2$ on the right-hand side:
\begin{align}
     \delta^2 Q|_{\rho_{\tilde{\bf m}}}
    &= \sum_{\bf m} c^{\tilde{\bf m}}_{\bf m}\left( \langle Q^2 \rangle_{\bf m} - \langle Q \rangle_{\bf m}^2 \right)
    \nonumber\\
    &\quad+\frac{1}{2} \sum_{{\bf m},{\bf m'}} c^{\tilde{\bf m}}_{\bf m}\, c^{\tilde{\bf m}}_{\bf m'}
    \left( \langle Q \rangle_{\bf m} - \langle Q \rangle_{\bf m'} \right)^2\nonumber\\
    &\equiv\mathbb{E}_{{\bf m}|\tilde{\bf m}} \left[ \delta^2 Q|_{\psi_{\bf m}} \right] + {\rm Var}_{{\bf m}|\tilde{\bf m}} \left[ \langle Q \rangle_{\bf m} \right].
\end{align}
Averaging over the random unitary gate set $\mathcal{U}$, measurement locations $\mathcal{M}$, and noisy measurement record $\tilde{\bf m}$ then yields the separation in Eq.~\eqref{eq:separation}.

For maximal readout noise, this separation acquires a sharp physical meaning. In this limit the assignment-error probability no longer depends on the ideal outcome vector $\bf m$, so $c^{\tilde{\bf m}}_{\bf m} = P_{\bf m}$: the noisy postselected ensemble $\rho_{\tilde{\bf m}}$ becomes identical to the noise-free, non-postselected ensemble $\rho_\varnothing$. The charge fluctuations in the maximally-noisy limit are therefore the noise-free fluctuations \emph{without} postselection and can be similarly split into two contributions:
\begin{align}
    \delta^2 Q|_{\rho_{\varnothing}} = \mathbb{E}_{{\bf m}} \left[ \delta^2 Q|_{\psi_{\bf m}} \right] + {\rm Var}_{\bf m} \left[ \langle Q \rangle_{\bf m} \right].
\end{align}
The second term, $\frac{1}{2} \sum_{{\bf m},{\bf m'}} P_{\bf m}\, P_{\bf m'} \left( \langle Q \rangle_{\bf m} - \langle Q \rangle_{\bf m'} \right)^2$, is classical in the sense that it is built entirely from first moments of the charge, $\langle Q \rangle_{\bf m}$. It is not, however, a fully benign quantity: it still depends explicitly on the measurement record. It is nonetheless essentially tomography-free, and previous work has shown that $\langle Q \rangle_{\bf m}$ can be estimated with high accuracy from the measurement record alone, without knowledge of the underlying quantum state, once the measurement rate exceeds $p_{\rm success} \gtrsim p_\#$~\cite{Vasseur2022learnability,Agrawal2024Observing_Learnability,ippoliti2023learnability}.

Deep in the charge-sharpening phase the quantum contribution to the charge fluctuations is essentially fully suppressed (for the sizes and timescales in our simulations, this corresponds roughly to $0.2 \lesssim p_m \leq 1$). The classical part from readout errors is then the only remaining source of fluctuations. Moreover, since this contribution is nearly uniform and $p_m$-independent throughout the sharpening phase, the finite-noise contribution takes the form $\overline{{\delta^2 Q}}_{\rm cl}(\eta)={\cal E}(\eta)\, \overline{{\delta^2 Q}}_{\rm cl}^{\rm max}$, where the maximal classical fluctuations at maximal noise ($\eta=1/2$) are
\begin{align}
\overline{{\delta^2 Q}}_{\rm cl}^{\rm max}&\equiv
{\mathbb E}_{\cal U,M}\left[ {\rm Var}_{\bf m} \langle Q \rangle_{\bf m} \right] \nonumber\\
&\approx
{\mathbb E}_{\cal U,M} \left[
\frac{1}{2 N_{\bf m}^2} \sum_{{\bf m},{\bf m'}} \left( \langle Q \rangle_{\bf m} - \langle Q \rangle_{\bf m'} \right)^2
\right].
\end{align}
The approximation reflects only finite sampling of trajectories (measurement records) in the simulation and becomes exact as $N_{\bf m}\to\infty$. 

The form for ${\cal E}(\eta)$ is simply the noise variance of a symmetric, uncorrelated noise channel. Indeed, the assignment error acts as an uncorrelated bit-flip channel at each measurement, superimposed on the intrinsic circuit dynamics: errors at different times and sites are independent of one another, but the \emph{ideal} measurement outcomes themselves are correlated and carry memory, since the probability of a given outcome for a qubit at time $t$ depends on the prior history of measurement outcomes as well as on the specific realization of the unitary gates. Through the weights $P_{\bf m}$ of the ideal outcomes, the noisy ensemble $\rho_{\tilde{\bf m}}$ therefore still retains correlated features tied to the full details of the specific circuit realization. Nevertheless, for large enough $p_m$ this memory effect is minimal, and the ensemble is essentially set by the most recent measurements. Taking the crude approximation $P_{\bf m}\sim 1/2^{M} \sim {\rm const.}$ (with $M$ the number of measurement sites) then recovers the uncorrelated-channel picture,
\begin{align}
    c_{\bf m}^{\tilde{\bf m}} \sim
    \eta^{\, d_{H}({\bf m},\tilde{\bf m})}
    (1-\eta)^{ M - d_{H}({\bf m},\tilde{\bf m})},
\end{align}
where $d_{H}$ is the Hamming distance between the two bit strings (the number of mismatches). This gives $\overline{\rm Var}_{{\bf m}|\tilde{\bf m}} \left[ \langle Q \rangle_{\bf m} \right] \propto \eta(1-\eta)$. Since $\overline{{\delta^2 Q}}_{\rm cl}^{\rm max}$ is defined as the classical fluctuation at $\eta=1/2$, the normalization $\mathcal{E}(1/2)=1$ is automatically fulfilled for ${\cal E}(\eta) = 4\eta(1-\eta)$. Our simulation data confirm this exact form (Fig.~\ref{fig:4}), particularly at large measurement rates, where the derivation's underlying assumption of uncorrelated channel is best justified.

\end{document}